\documentclass{elsart}
\usepackage{epsfig}
\begin{document}
       \runauthor{Ohnuki,Snowden-Ifft, Martoff}
\begin{frontmatter}
    \title{Measurement of Carbon Disulfide Anion Diffusion in a TPC}
\author[Occidental]{Tohru Ohnuki\thanksref{email}},
\author[Occidental]{Daniel P. Snowden-Ifft}
\author[Temple]{C. J. Martoff}
\address[Occidental]{Department of Physics, Occidental College, 1600 Campus Road, Los 
Angeles 90041-3314, USA}
\address[Temple]{Department of Physics, Temple University, 1900 N. 
13-th St., Philadelphia, PA 19122-6082, USA}
\thanks[email]{Email: ohnukit@oxy.edu}
\begin{abstract}
A Negative Ion Time Projection Chamber was used to measure the field
dependence of lateral and longitudinal diffusion for CS$_2$ anions
drifting in mixtures of CS$_2$ and Ar at 40 Torr.  Ion drift
velocities and limits on the capture distance for electrons as a
function of field and gas mixture are also reported.
\end{abstract}
  \begin{keyword}
    carbon disulfide, CS$_{2}$, diffusion, mobility, TPC\\
    PACS Classification: 29.40.Cs, 29.40.Gx, 51.50.+v
\end{keyword}
\end{frontmatter}

\section{Introduction}
The Directional Recoil Identification From Tracks (DRIFT) detector has
recently been proposed \cite{driftprd} to search for dark matter. 
This detector is unique in that it drifts negative ions, instead of
electrons, in a time projection chamber (TPC).  A detailed description
of the operation, motivation, and other uses of a negative ion TPC
(NITPC) can be found in \cite{martoff}.  Briefly, an electronegative
component in the gas captures ionized electrons forming negative ions. 
These anions then drift to the anode wires where, provided the
electronegative component allows it, the anions are ionized and
normal electron avalanche occurs.  The motivation for such an
arrangement is that it allows transport of charge to the anode with
the minimum possible diffusion.  Using a proportional chamber, Crane
showed that CS$_{2}$ has the desired electronegative properties
\cite{Crane}.  Using a single wire drift chamber of unusual design,
Martoff et al.  \cite{martoff} measured the drift velocity and lateral
diffusion of CS$_{2}$ ions in two different gas mixtures.  In this
paper we used a NITPC to measure the drift velocity, lateral diffusion
and longitudinal diffusion in a variety of CS$_{2}$-Ar gas mixtures. 
These measurements allowed limits to be set on the electron capture
distance in these mixtures.

\section{Theory}\label{sec:theo}
The diffusion of charged particles being drifted through a gas has
been parameterized by an expression of the form \cite{Rolandi&Blum}:
\begin{equation}\label{eqn:canonical}
    \sigma^{2} =\frac{4 \varepsilon_{\mathrm{k}} L}{3 e E}
\end{equation}
where $L$ is the drift distance, $E$ is the drift field and
$\varepsilon_{\mathrm{k}}$ is the characteristic (average) energy of
the electron or ion.  For electrons, $\varepsilon_{\mathrm{k}}$ varies
from thermal ($\sim k_{\mathrm{B}}T$) at low drift fields to several
eV at higher $E/p$.  The nonlinear variation of
$\varepsilon_{\mathrm{k}}$ for electrons arises from the mass mismatch
between electrons and gas atoms which prevents elastic electron--atom
collisions from efficiently thermalizing the electron energy gained
between collisions \cite{Sauli_Bible}.  Ions, on the other hand, have
masses comparable to the gas atoms and hence would be expected to
remain well thermalized even if the energy gained between collisions
became comparable with $\sim k_{\mathrm{B}}T$.  Thus
$\varepsilon_{\mathrm{k}}$ for ions should remain constant and one
should be able to reduce the diffusion of ions (as 1/$\sqrt{E}$) to
much lower values than for electrons.  The NITPC concept relies
critically on this hypothesis.  The purpose of the present work was to
test it in detail.

\section{Experimental Method}
\subsection{Apparatus}
\begin{figure}[ht]
    \centerline{\epsfig{file=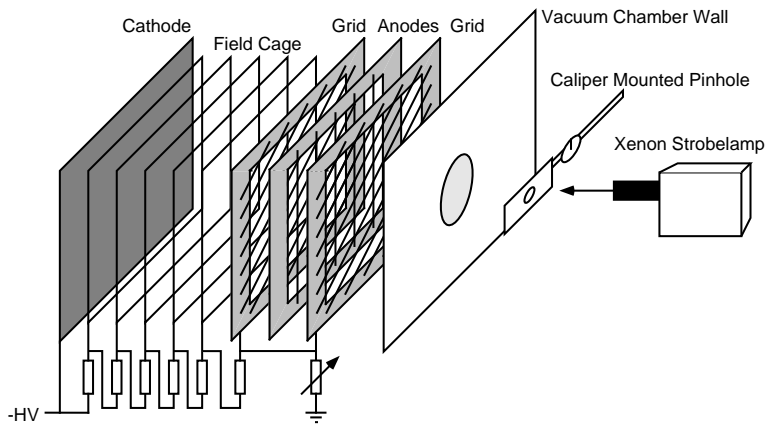,width=4in}}
    \caption{TPC in Situ (not to scale)}\label{fig:app}
\end{figure}
The NITPC used in this experiment consisted of a drift region attached
to a multi-wire proportional chamber (MWPC) (Figure \ref{fig:app}). 
The drift region was composed of a solid copper cathode and field
cage.  The field cage consisted of 4 loops of 500 $\mu$m wire
supported by acrylic posts and connected by a ladder of precision 10
M$\Omega$ resistors.  The lateral dimensions of the field cage were 15
cm by 20 cm.  The MWPC was made of three wire planes: two grid planes
sandwiching an anode plane.  The grid and anode wires were 100 $\mu$m
and 20 $\mu$m gold coated tungsten respectively.  Anode wire spacing
was 2 mm, grid wire spacing was also 2 mm but at 45 degrees to the
anodes and the gap was 1.1 cm.  The drift distance, from the solid
copper cathode to the nearest grid plane, was 10 cm.

A single negative high voltage supply connected at the cathode was
used to set the drift and MWPC potentials via the field cage
resistance and a variable resistance between anode and grid.  During
operation, the MWPC voltage was kept as high as possible to yield the
greatest gas gain and to insure 100\% transparency \cite{Rolandi&Blum}.
 
The detector was mounted with wire planes vertical in an evacuable
chamber.  Gas mixtures were prepared in the chamber by introducing
gases one by one through a manifold.  CS$_{2}$ vapor was evolved from
a liquid source by the low pressure in the chamber.  All experiments
were done at a pressure of 40.0 Torr.  The chamber pressure was
measured with a capacitive manometer allowing accurate (0.1 Torr)
determination of pressure and mixture.  Before each run, the chamber
was evacuated to $\sim$ 50 mTorr measured using a Convectron gauge, then
backfilled with the desired gas mixture and pressure.  A gas filling
would be used for several hours, during which time the pressure rise
was about a percent.

The diffusion measurements were made using photoelectrons generated
from the solid copper cathode in the following way.  Near the NITPC
was a port containing a fused silica window.  A UV flashlamp projected
light through a movable 200 $\mu$m aperture, through the silica
window, through the grid and anode wire planes, and onto the cathode. 
The projected spot on the cathode was $\sim$ 1.3 mm by $\sim$ 2 mm
measured using photographic film.  To maintain the photoelectron
yield, the copper cathode was cleaned with abrasive every three to
five runs.  A photodiode viewing the flashlamp directly provided a
start signal for drift time measurements.

The anode wires were connected to Amptek A-250 charge sensitive
pre-amplifiers which were housed inside the chamber to reduce noise
pickup.  The outputs of the Ampteks were amplified and shaped by Ortec
855 spectroscopy amplifiers.  These signals were digitized and stored
on a 20 MHz digital oscilloscope.  The scope used the photodiode
signal as a trigger.  Waveform averaging was used to improve signal to
noise.

\subsection{Measurement}
To measure the lateral diffusion, the maximum pulse heights on two
adjacent wires were measured as the aperture was moved laterally
through 40, 125 $\mu$m steps.  Pulse height maxima were measured on
waveforms averaged over either 128 or 256 individual flashlamp shots. 
Space charge should not affect the results of this experiment.  The
number of photoelectrons produced was always less than $\sim$ 50 per
flash.  The lateral diffusion of the anions was always greater than
0.7 mm.  Since the gain was low, $\sim$ 3000, the size of the
avalanche was $\sim$ 0.05 mm \cite{Rolandi&Blum}.  In those cases
where two avalanches overlapped, the reduction in gain for the second
avalanche was less than 1\% \cite{Rolandi&Blum}.  In addition to
these theoretical arguments, the gain linearity was tested by
confirming that a $\times$10 reduction in flash energy resulted in an
order of magnitude reduction in pulse height.
\begin{figure}[b]
    \centerline{\epsfig{file=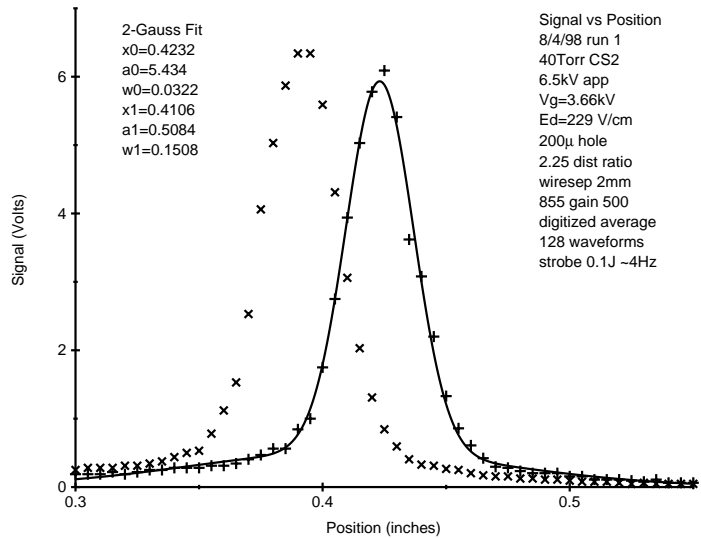,width=3.75in}}
    \caption{Two--Gaussian Fit to Typical Data Set}\label{fig:2G}
\end{figure}

Figure \ref{fig:2G} shows a set of measurements, with aperture
position on the horizontal axis and pulse height of the averaged
waveforms on the vertical axis.  The data for each peak was fit to a
function that allowed deconvolution of the deposited charge from a
known electronic artifact.  The distance between the two peaks
represented the 2 mm wire spacing and was used to normalize the
lateral measurement.  The +'s and $\times$'s in Figure \ref{fig:2G}
are the data for each wire, the line is the fitted function.

The longitudinal diffusion was determined by measuring the pulse width
in time on a single wire.  To convert this measurement to distance,
the drift speed as a function of drift field was determined for each
gas mixture.  This was done by measuring the delay time between the
light pulse (photoelectron generation) and the peak of the anode
signal.  Though the majority of the delay was due to the time ions
spent in the drift field, a correction was made for the delay in the
MWPC.

In order to minimize variations due to changing chamber conditions, a
set of runs at different drift fields would be done with a given gas
mixture.  Additionally, all of the above measurements (lateral and
longitudinal diffusion and drift velocity) were done concurrently at a
given drift field.  The delay time and pulse time-width were both
measured at the aperture position corresponding to maximum pulse
height for the wire under test.  To reduce the effects of gas aging, a
gas mixture was kept for only one set of runs, and was flushed at the
end of the day.

\section{Results and Discussion}
There are a number of contributions to the {\em measured} $\sigma$. 
The main ones are the finite spot size, geometry of the chamber,
capture distance, and the diffusion itself(given by Equation
\ref{eqn:canonical}).  Since the gap distance in the MWPC was so small
and the field so large the diffusion there was assumed to be minimal. 
The various contributions can be assumed to be uncorrelated and hence
to add in quadrature:
\begin{equation}\label{eqn:difsum}
    \sigma_{\mathrm{measured}}^{2} = \sigma_{\mathrm{spot}}^{2} + 
    \sigma_{\mathrm{geometry}}^{2} + 
    \sigma_{\mathrm{capture}}^{2} + \frac{2 k_{\mathrm{B}}
T_{\mathrm{eff}} L}{e} \frac{1}{E}
\end{equation}
\noindent where $\varepsilon_{\mathrm{k}}$ in Equation
\ref{eqn:canonical} has been characterized in terms of an effective
temperature $T_{\mathrm{eff}}$.  The NITPC hypothesis is that 
$T_{\mathrm{eff}}$ will
remain constant and close to room temperature up to very high drift
fields.

For the lateral measurements, the $\sigma_{\mathrm{spot}}$ was significant
while for the longitudinal measurements it was not due to the short
duration of the pulse.  In the lateral measurements
$\sigma_{\mathrm{geometry}}$ arises from the 2 mm wire spacing and is
significant.  For the longitudinal measurements $\sigma_{\mathrm{geometry}}$
arises from the different pathlengths ions can take to get to the
anode wires and is also significant.  None of these is a function of
the field.
  
Since $\sigma_{\mathrm{capture}}$ is expected to increase with increasing
field, a linear relationship between $\sigma_{\mathrm{measured}}^{2}$ 
and $1/E$
supports the assumption that $\sigma_{\mathrm{capture}}$ is slowly varying or
constant.  In that case, $T_{\mathrm{eff}}$ can be deduced from the slope.  The
capture distance can be deduced from the intercept if the geometry and
spot size contributions are known or estimated, since according to
Equation \ref{eqn:difsum}
\begin{equation}\label {eqn:intercept}
    \sigma_{\mathrm{capture}}^{2} = \sigma_{\mathrm{intercept}}^{2} - 
    \sigma_{\mathrm{geometry}}^{2} - 
    \sigma_{\mathrm{spot}}^{2}
\end{equation}

\subsection{Lateral Diffusion}
\begin{figure}[tb]
    \centerline{\epsfig{file=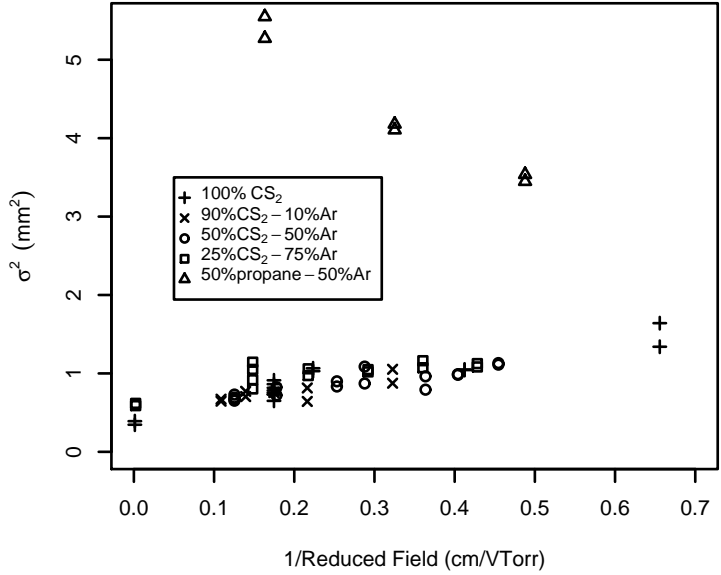,width=4.0in}}
    \caption{Lateral Diffusion for Various Gas Mixtures}\label{NMD}
\end{figure}

The main lateral diffusion results are summarized in Figure \ref{NMD}
and Table \ref{difrel}.  The plot shows the lateral diffusion squared
($\sigma^{2}$) versus the inverse of the drift field.
\begin{table}[tb]
\caption{Lateral Diffusion Results}\label {difrel}
\vspace{1ex}
\footnotesize
\centerline{
\begin{tabular}[]{lcccc}
    \hline
    Gas Mixture          & Slope                   & Temperature   & Y-Intercept             & $\sigma_{capture}$\\
    \hline \hline
    100\%CS$_{2}$        & 0.16 $\pm$0.02 Vmm/Torr & 360 $\pm$40 K & 0.50 $\pm$0.05
    mm$^{2}$ & $<$0.4 mm\\
    90\%CS$_{2}$-10\%Ar & 0.13 $\pm$0.03 Vmm/Torr & 300 $\pm$80 K & 0.52 $\pm$0.07
    mm$^{2}$ & $<$0.4 mm\\
    50\%CS$_{2}$-50\%Ar & 0.11 $\pm$0.02 Vmm/Torr & 260 $\pm$40 K & 0.56 $\pm$0.05
    mm$^{2}$ & $<$0.5 mm\\
    25\%CS$_{2}$-75\%Ar & 0.10 $\pm$0.02 Vmm/Torr & 240 $\pm$50 K & 0.74 $\pm$0.06
    mm$^{2}$ & $<$0.6 mm\\
    \hline
\end{tabular}
}
\normalsize
\end{table}
On the graph, the marks represent the data taken for different gas
mixtures at a variety of drift fields.  The triangles are electrons
drifted in 50\% propane-50\%argon for reference.  The points very
close to the origin are from experiments where the drift region was
removed and the MWPC used alone with a copper cathode in place of one
of the grid planes.  The copper cathode was in approximately the same
position relative to the light source for both the MWPC and NITPC
measurements so that the spot size was constant.  The reason that
several diffusion values appear for each drift field is that two wires
were used during the measurement.  Since the uncertainty in the
Gaussian fits for the points is much smaller than the marks, the
separation between the points at identical fields represents a
systematic error.  This is believed to be due to the modulation of UV
light intensity by the grid and anode wires.

Clearly, the electron data (open triangles in Figure \ref{NMD}) do not
obey Equation (2), but all of the ion data do.  The ion data were fit
to lines from which $T_{\mathrm{eff}}$ and the intercept were extracted.  As
can be seen in Table \ref{difrel}, all of the temperatures are
consistent with thermal diffusion with perhaps a hint that the lateral
diffusion is decreasing with increasing concentration of Ar.
 
The fact that the high field points lie on the general trend of the
data is important in that it shows that electrons are captured even in
very large E/p.  For the 100\% CS$_{2}$ data the high field was 83
V/cmTorr while for the 25\%\, CS$_{2}$ - 75\%\,Ar data it was 45
V/cmTorr.  The intercept data shown in Table \ref{difrel} and Equation
\ref{eqn:intercept} can now be used to figure out $\sigma_{\mathrm{capture}}$. 
Given the crude nature of our photographic measurement 
$\sigma_{\mathrm{spot}}$
could not be estimated with any reliability.  However,
$\sigma_{\mathrm{geometry}} = 2 mm / \sqrt{12}$ which allows to us to place the
upper limits on $\sigma_{\mathrm{capture}}$ shown in Table \ref{difrel}.

\subsection{Mobility}
\begin{figure}[t]
    \centerline{\epsfig{file=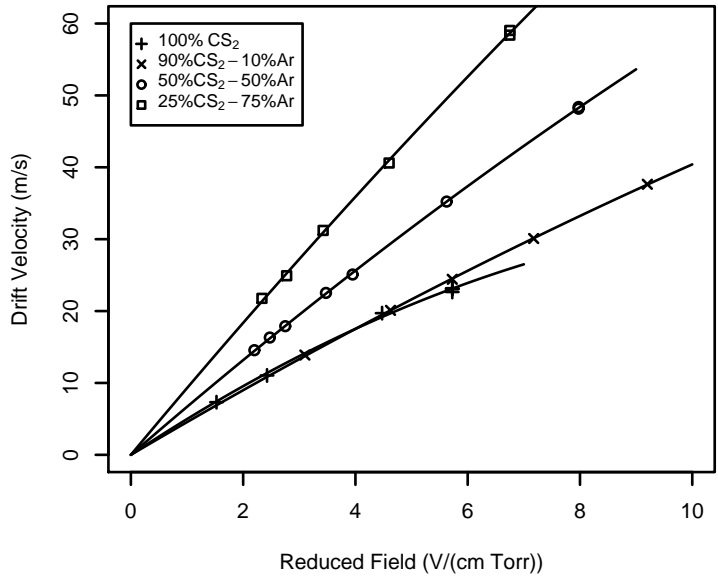,width=4.0in}}
    \caption{Drift Velocity vs. Field}\label{vd}
\end{figure}

Figure \ref{vd} shows the drift velocity data for the gas mixtures
plotted against the reduced field.  The data from each of the gas
mixtures was fitted to $v=m(E/p) + n(E/p)^{2}$.

\begin{table}[t]
    \caption{Mobility Coefficients}\label{mob}
\vspace{1ex}
\centerline{    
\begin{tabular}[]{lcc}
    \hline
    Gas Mixture & m & n\\
    \hline \hline
    100\%CS$_{2}$        & 5.2  $\pm$0.2  & -0.20  $\pm$0.03  \\
    90\%CS$_{2}$-10\%Ar & 4.6  $\pm$0.03 & -0.057 $\pm$0.004 \\
    50\%CS$_{2}$-50\%Ar & 6.77 $\pm$0.03 & -0.090 $\pm$0.005 \\
    25\%CS$_{2}$-75\%Ar & 9.4  $\pm$0.1  & -0.10  $\pm$0.02  \\
    \hline
\end{tabular}}
\end{table}
Table \ref{mob} shows the drift velocity coefficients for the
different gas mixtures.  The drift velocity increases with increasing
concentration of argon.  This is not unexpected as an Ar atom is much
smaller than a CS$_{2}$ molecule and therefore the mean free path of
the CS$_{2}$ ions is much larger, the mean time to collision is
increased, thus yielding a higher drift velocity.  In addition the
larger mass of the CS$_{2}$ molecule means that it will preferentially
scatter off of the Ar atoms in the forward direction.

\subsection{Longitudinal Diffusion}
\begin{figure}[b]
    \centerline{\epsfig{file=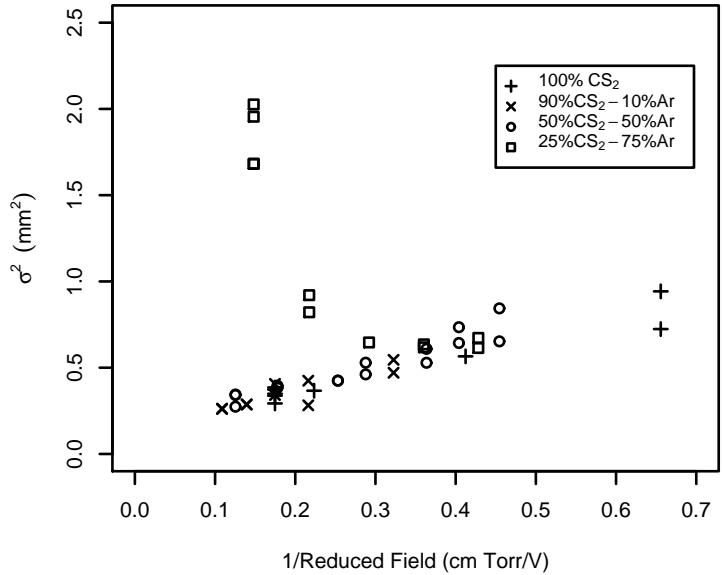,width=4.0in}}
    \caption{Longitudinal Diffusion}\label{NMT}
\end{figure}

The longitudinal diffusion results are shown in Figure \ref{NMT} and
Table \ref{zrel}.  As in the lateral case, the rms diffusion squared
is plotted against the inverse of the reduced drift field.

As before, the marks represent the data for the various gas mixtures. 
All of the data look linear with the obvious exception of the
25\%CS$_{2}$-75\%Ar mixture.  The linear datasets were fit to
straight lines, giving the $T_{\mathrm{eff}}$ values and intercepts shown in
Table \ref{zrel}.  Notice that in this case the trend is for the
temperature to increase with increasing concentration of Ar.
\begin{table}[t]
    \caption{Longitudinal Diffusion Results}\label {zrel}
\vspace{1ex}
\footnotesize
\centerline{  
\begin{tabular}[]{lcccc}
    \hline
    Gas Mixture          & Slope                   & Temperature   & Y-Intercept             & $\sigma_{capture}$\\
    \hline \hline
    100\%CS$_{2}$        & 0.10 $\pm$0.01 Vmm/Torr & 230 $\pm$20 K & 0.17 $\pm$0.02
    mm$^{2}$ & $<$0.4 mm\\
    90\%CS$_{2}$-10\%Ar & 0.11 $\pm$0.02 Vmm/Torr & 260 $\pm$40 K & 0.14 $\pm$0.04
    mm$^{2}$ & $<$0.4 mm\\
    50\%CS$_{2}$-50\%Ar & 0.13 $\pm$0.01 Vmm/Torr & 300 $\pm$20 K & 0.14 $\pm$0.04
    mm$^{2}$ & $<$0.4 mm\\
    \hline
\end{tabular}
}
\normalsize
\end{table}
 
Unlike the lateral case, $\sigma_{\mathrm{spot}}$ is small since photoelectrons
are created at the same time.  Also unlike the lateral case,
$\sigma_{\mathrm{geometry}}$ is difficult to calculate since it arises from the
different path lengths traveled by the ions.  It is probably
significant on the scale of these measurements (i.e. sub-mm). 
Therefore we can again only infer an upper limit on the capture
distance, shown in Table \ref{zrel} for the three measurements which
indicated a constant capture distance.

Given the discussion accompanying Equation \ref{eqn:difsum} the
interpretation of the 25\%CS$_{2}$-75\%Ar data is that
$\sigma_{\mathrm{capture}}$ is a stronger function of the field for this
mixture.  This interpretation is bolstered by a measurement of a
10\%CS$_{2}$-90\%Ar mixture in which some of the electrons drifted
directly to the anode wires without being captured by the CS$_{2}$. 
This was seen as a ``direct'' pulse, arriving within several
microseconds after the light pulse instead of the several milliseconds
that the ions take.

\subsection{Trends}
The lateral and longitudinal ion diffusion data are generally
consistent with diffusion at thermal ion energies, with the exceptions
noted above (25\%CS$_{2}$-75\%Ar mixtures).

If a further trend exists it is for the lateral diffusion to decrease
with increasing Ar concentration while the opposite occurs for the
longitudinal diffusion.  Removing the systematic error discussed above
may help to resolve these trends.  Both of these deviations from
Equation \ref{eqn:canonical} can be understood from the fact that
CS$_{2}$ is much heavier than Ar and therefore tends to preserve its
direction of motion after a collision with Ar, violating one of the
assumptions underlying Equation \ref{eqn:canonical}.

The $\sigma_{\mathrm{capture}}$ limits are harder to understand.  For gas
mixtures 50\% Ar or less the lateral and longitudinal data both
indicate a constant value with respect to field and gas mixture.  One
would expect each of these to increase with increasing Ar
concentration but this increase may be masked by the constant terms in
Equation \ref{eqn:intercept}.  At 75\% Ar the lateral and longitudinal
data are very different.  The lateral data seem to indicate an
elevated though field independent $\sigma_{\mathrm{capture}}$ while the
longitudinal data clearly indicate a field dependent
$\sigma_{\mathrm{capture}}$.  At 90\% Ar the prompt arrival of electrons is
evidence for a $\sigma_{\mathrm{capture}}$ larger than 10 cm.

\section{Conclusion} 

The lateral and longitudinal diffusion of CS$_{2}$ anions were
systematically studied as a function of field and gas mixture.  While
mostly consistent with a thermal model certain trends were identified
which need to be explored in more detail.  The drift velocity of
CS$_{2}$ anions and limits on the capture distance for producing those
anions were also measured.  These measurements are critical for the
operation of the DRIFT detector and for future uses for NITPCs.  The
results show that the NITPC concept will permit detectors such as
DRIFT\cite{driftprd} to achieve submillimeter track diffusion for
drift distances of the order of a meter, over a useful range of gas
mixtures and drift fields.

\section{Acknowledgments}

Funding for this project was provided by:
\begin{itemize}
\item[] Academic Student Project Award of Occidental College (spring 98, fall
98)
\item[] Ford-Anderson Summer Research Grant (summer 98)
\item[] Research Corporation Grant No. CC--4512 (summer 98)
\end {itemize}

\section{References}

\bibliography{cs2diff}

\begin{thebibliography}{1}

\bibitem{driftprd}
Daniel~P. Snowden-Ifft, C.~Jeff Martoff, and Juan~M. Burwell.
\newblock Low pressure negative ion drift chamber for dark matter search.
\newblock {\em Accepted for publication in Physical Review D}, 2000.

\bibitem{martoff}
C.~Jeff Martoff, Daniel~P. Snowden-Ifft, Tohru Ohnuki, Neil J.~C. Spooner, and
  Matthew Lehner.
\newblock Supressing drift chamber diffusion without magnetic field.
\newblock {\em Nucl. Instr. \& Meth. in Phys. Res. A}, 440:335, 2000.

\bibitem{Crane}
H.~R. Crane.
\newblock {CO$_2$-CS$_2$} geiger counter.
\newblock {\em Rev. Sci. Inst.}, 32:953, 1961.

\bibitem{Rolandi&Blum}
Luigi Rolandi and Walter Blum.
\newblock {\em Particle Detection with Drift Chambers}.
\newblock Springer-Verlag, Berlin, Germany, 1994.

\bibitem{Sauli_Bible}
F.~Sauli.
\newblock {\em Experimental Techniques in High Energy Physics}, page~81.
\newblock Addison-Wesley Publishing, 1987.

\end{thebibliography}
\bibliographystyle{unsrt}



\end{document}